\documentclass[aps,prd,twocolumn,amsmath,groupedaddress]{revtex4}
\usepackage{epsfig}

\def\al{\alpha}

\def\de{\delta}
\def\ep{\epsilon}

\def\ka{\kappa}
\def\la{\lambda}

\def\si{\sigma}
\def\vs{\varsigma}

\def\ch{\chi}

\def\La{\Lambda}
\def\Si{\Sigma}

\def\Om{\Omega}

\def\cl{{\cal L}}

\def\frac#1#2{\textstyle{{{#1} \over {#2}}}}

\def\prt{\partial}
\def\vev#1{\langle {#1}\rangle}

\def\half{{\textstyle{1\over 2}}}
\def\lsim{\mathrel{\rlap{\lower4pt\hbox{\hskip1pt$\sim$}}
    \raise1pt\hbox{$<$}}}
\def\gsim{\mathrel{\rlap{\lower4pt\hbox{\hskip1pt$\sim$}}
    \raise1pt\hbox{$>$}}}

\def\etal {{\it et al.}}

\newcommand{\beq}{\begin{equation}}
\newcommand{\eeq}{\end{equation}}
\newcommand{\bea}{\begin{eqnarray}}
\newcommand{\eea}{\end{eqnarray}}
\newcommand{\bse}{\begin{subequations}}
\newcommand{\ese}{\end{subequations}}
\newcommand{\rf}[1]{(\ref{#1})}

\def\sqr#1#2{{\vcenter{\vbox{\hrule height.#2pt
         \hbox{\vrule width.#2pt height#1pt \kern#1pt
         \vrule width.#2pt}
         \hrule height.#2pt}}}}

\def\kf{\hat k_F}
\def\kaf{\hat k_{AF}}
\def\kfd#1{k_{F}^{(#1)}}
\def\kafd#1{k_{AF}^{(#1)}}
\def\klm#1#2#3{k^{(#1)}_{(#2)#3}}
\def\sylm#1#2{\phantom{}_{#1}Y_{#2}}
\def\mbf#1{\mbox{\boldmath$#1$}}

\begin{document}

\title{Lorentz-Violating Electrodynamics 
and the Cosmic Microwave Background}

\author{V.\ Alan Kosteleck\'y$^1$ and Matthew Mewes$^2$}
\address{$^1$Physics Department, Indiana University, 
Bloomington, IN 47405, U.S.A.\\
$^2$Physics Department, Marquette University,
Milwaukee, WI 53201, U.S.A.}

\date{IUHET 504, January 2007; 
accepted for publication in \it Physical Review Letters \rm} 

\begin{abstract}
{\noindent
Possible Lorentz-violating effects 
in the cosmic microwave background are studied.
We provide a systematic classification 
of renormalizable and nonrenormalizable operators 
for Lorentz violation in electrodynamics
and use polarimetric observations  
to search for the associated violations. 
}
\end{abstract}

 
\maketitle 

Relativity has been confirmed to a high degree 
of precision by many experiments over the past century.
Recent years have seen renewed interest in sensitive tests 
of relativity following the realization 
that tiny violations of Lorentz symmetry,
which forms the basis of both Special and General Relativity,
can arise in theories that attempt to unify all known forces
\cite{ks}.
While contemporary searches for Lorentz violations involve
many types of matter and energy,
the properties of light have traditionally been the primary focus.
Today,
searches for dominant relativity-violating effects
involving photons include modern versions of
the classic Michelson-Morley and Kennedy-Thorndike experiments
\cite{lipa,muller,stanwix,herrmann,antonini,stanwix2}
and analyses of polarized light from distant astrophysical sources
\cite{cfj,km}.
The latter take advantage of the extreme propagation times 
over which tiny effects can accumulate,
and they yield sensitivities comparable 
to those achieved with matter 
\cite{bear,wolf,heckel}.
The cosmic microwave background (CMB),
which is the oldest untainted radiation available to observation,
offers a unique opportunity for Lorentz-violation searches
involving photons.
In this work,
we introduce a systematic classification 
of coefficients for Lorentz violation at all orders,
develop theoretical tools to extract sensitivity
from polarimetric observations of the CMB,
and analyse observational data
to obtain first measurements of various 
relativity-violating effects. 

At attainable scales,
Lorentz violations are described by 
the Standard-Model Extension (SME)
\cite{sme}.
The SME is an effective field theory
that serves as the general theoretical basis 
for experimental searches, 
including ones with light.
It categorizes the type of Lorentz violation
by the mass dimension $d$ of the
corresponding operator in the Lagrange density,
which offers a simple measure of their expected size
\cite{kp}.
Existing studies of the SME photon sector 
primarily focus on operators of renormalizable dimension $d\leq 4$,
but here we consider terms with arbitrary $d$
that preserve the usual U(1) and spacetime-translation symmetries
and hence conserve charge, energy, and momentum.
Some calculation reveals that in this case 
the photon sector of the SME Lagrange density takes the form 
\bea
\cl &=&  -\frac 1 4 F_{\mu\nu}F^{\mu\nu}
+\frac 1 2 \ep^{\ka\la\mu\nu}A_\la (\kaf)_\ka F_{\mu\nu}
\nonumber \\  
&& 
- \frac 1 4 F_{\ka\la} (\kf)^{\ka\la\mu\nu} F_{\mu\nu} ,
\label{Lagrange density}
\eea
where $A_\mu$ is the electromagnetic 4-potential
and $F_{\mu\nu}$ is the field-strength tensor.
The first term in $\cl$ is conventional Maxwell electrodynamics,
while the other terms violate Lorentz symmetry.
The quantities $(\kaf)_\ka$ and $(\kf)^{\ka\la\mu\nu}$ 
are polynomials in the 4-momentum operator $p_\mu=i\prt_\mu$
given by 
\begin{align}
(\kaf)_\ka &=\hspace{-3pt}\sum_{d {\rm ~odd}} 
{(\kafd{d})_\ka}^{{\al_1}\ldots{\al_{(d-3)}}} 
\prt_{\al_1}\ldots\prt_{\al_{(d-3)}} \ ,
\nonumber\\
(\kf)^{\ka\la\mu\nu} &= \hspace{-5pt}\sum_{d {\rm ~even}} 
(\kfd{d})^{\ka\la\mu\nu{\al_1}\ldots{\al_{(d-4)}}} 
\prt_{\al_1}\ldots\prt_{\al_{(d-4)}} \ ,
\label{kfs}
\end{align}
where $\kafd{d}$, $\kfd{d}$ 
are constant coefficients for Lorentz violation 
of dimension $4-d$.
The coefficients $\kafd{d}$ violate CPT symmetry,
while the coefficients $\kfd{d}$ preserve it.
If these coefficients emerge from spontaneous breaking,
the associated Nambu-Goldstone modes might play the
role of the photon 
\cite{ng},
but this issue is secondary and disregarded here. 
Note that relaxing U(1) invariance
would introduce a $d=2$ photon-mass term,
among other effects.

The operators in Eqs.\ \rf{kfs}
produce changes in the properties of electromagnetic radiation.
The plane-wave solutions to the equations of motion 
obtained from Eq\ \rf{Lagrange density}
reveal that in the presence of Lorentz violation
light propagating in empty space
can be viewed as a superposition of two modes 
differing in polarization and velocity. 
The difference in phase velocity between the modes 
causes a shift in the relative phase 
between the two modes during propagation,
which alters the superposition
and thereby produces cosmic birefringence.
For each type of operator causing birefringence,
the size of the effect is governed by 
the associated coefficient for Lorentz violation 
multiplied by a factor of $E^{d-3} t$,
where $E$ is the photon energy and $t$ is the propagation time.
For cosmological sources this factor can become very large,
providing extreme sensitivity to minuscule violations
of Lorentz invariance.

\begin{table*}
\renewcommand{\arraystretch}{1.2}
\begin{tabular*}{\textwidth}{@{\extracolsep{\fill}}c|ccccc}
\hline
$d$ & 3 & 4 & 5 & 6 & 7 \\
\hline
$H_0\int (E/E_{\mbox{\scriptsize ob}})^{d-3} dt$ & 
0.95 & 2.7 & 40 & $8.6\times10^3$ & $4.9\times10^6$\\
Estimated sensitivity\ (GeV$^{4-d}$) 
& $10^{-42}$ & $10^{-29}$ & $10^{-18}$ & $10^{-7}$ & $10^{3}$\\
B03 sensitivity\ (GeV$^{4-d}$) 
& $10^{-42}$ & $10^{-30}$ & $10^{-19}$ & $10^{-9}$ & - \\
\hline
\end{tabular*}
\caption{
CMB sensitivities to Lorentz-violating operators 
of dimension $d$.
The first row lists numerical values of the energy integral 
in terms of $E_{\mbox{\scriptsize ob}}$ and 
the Hubble constant $H_0=71$ km/s/Mpc.
The second row gives the estimated sensitivity 
to the corresponding coefficient for Lorentz violation. 
The third row lists approximate sensitivities 
we obtain by comparison with B03 data.
The cosmological parameters adopted in this work are
$z_{\mbox{\scriptsize CMB}}=1100$,
$\Om_m=0.27$,
$\Om_\La=0.73$,
$\Om_r=0.015$.
}
\end{table*}

The CMB radiation is now known to be partially polarized
\cite{dasi_pol,capmap_pol,cbi_pol,wmap_pol,boom_pol}
and has propagated for approximately 14 billion years,
so even minuscule Lorentz violations could alter its polarization
in a detectable way 
\cite{feng,cmb}.
For CMB radiation, taking 
the observed photon energy as
$E_{\mbox{\scriptsize ob}} \sim 10^{-13}$ GeV
and the propagation time as
$t \sim 10^{10}$yr $\sim 10^{42} $ GeV$^{-1}$,
we obtain a crude estimated sensitivity of parts in 
$10^{81-13d}$ GeV$^{4-d}$
to dimension-$d$ coefficients for Lorentz violation.
Since the sensitivity to relativity violations 
grows roughly as $E^{d-3}$,
higher photon energies generally lead to higher sensitivities.
One might therefore expect 
studies of the lower-energy microwaves in the CMB
to yield lesser sensitivities
than prior searches for birefringence from SME operators with $d=4$
performed using near-optical emissions from distant galaxies
and gamma rays from gamma-ray bursts
\cite{km}.
However,
a significant advantage arises from the cosmological redshift.
Much of the polarization change occurs 
shortly after the CMB was produced,
when the Universe was much hotter and the photons
were approximately 1000 times more energetic.
This implies that studying the CMB 
is effectively equivalent to an optical test
with a time scale set by some fraction of the Hubble time.
Indeed,
explicitly integrating the CMB energy 
from the time of last scattering to the present
reveals that for operators with $d > 5$
the effective sensitivity to Lorentz violation 
is well approximated by parts in 
$10^{67-10d}/(d-5)$ GeV$^{4-d}$,
a substantial improvement over the crude estimate.
Table I provides numerical values of the integral 
and estimated sensitivities.

\begin{figure*}
\vskip -12pt
\begin{center}
\centerline{\psfig{figure=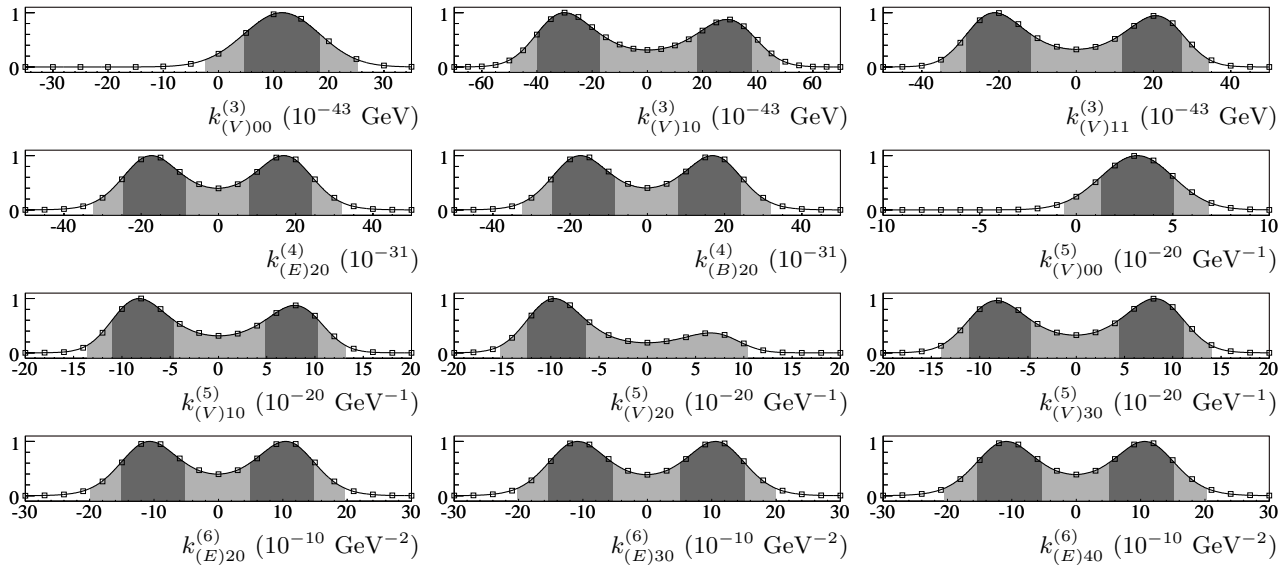,width=0.95\hsize}}
\vskip -14pt
\caption{
Sample plots of likelihood versus 
values of coefficients for Lorentz violation.
For each listed coefficient,
the boxes indicate numerically calculated values,
and the curve is a smooth extrapolation through them.
Dark-gray and light-gray regions represent 
the 68\% and 95\% confidence levels,
respectively.
}
\end{center}
\vskip -10pt
\end{figure*}

To study the implications of Lorentz violation
for the observed CMB polarization,
we must first understand the effects 
on the Stokes parameters $Q$, $U$, $V$
and the corresponding Stokes vector
$\mbf s = (s^1,s^2,s^3)^T = (Q,U,V)^T$.
The birefringence induced from Eq.\ \rf{Lagrange density}
causes the Stokes vector $\mbf s$ characterizing 
the net polarization of the light to rotate about an axis 
given by the Stokes vector $\mbf\vs$ for the faster mode
\cite{km}.
The angle of rotation of $\mbf s$ is the change in relative phase.
For a convenient normalization of $\mbf\vs$,
the differential rotation is given at leading order by
\beq
d\mbf s/dt = 2 E \mbf \vs\times \mbf s 
\equiv -i \Si\cdot \mbf s\ ,
\label{stokesrot}
\eeq
where $\Si$ is a matrix.
The components of $\mbf\vs$ and hence of $\Si$
control completely the polarization change
as light propagates from a distant source to Earth,
and they depend on coefficients for Lorentz violation,
the photon frequency, and the propagation direction.

For given values of the coefficients for Lorentz violation,
the change in polarization as light propagates 
is determined by integration of Eq.\ \rf{stokesrot}
from emission to detection.
For the CMB radiation,
the integration must be done for each point on the sky,
and two issues must be addressed.
The first is the cosmological redshift,
which leads to decreasing photon frequencies 
and consequent changes in the rotation axis
$\mbf\vs$ as the light propagates.
This typically makes analytical considerations sufficiently
challenging that numerical integration is needed.
The second issue 
involves the tensor nature of the Stokes parameters
and the whole-sky nature of the CMB.
For light propagating inward over the sphere of the sky,
the Stokes parameters $s^1$ and $s^2$
are components of a symmetric 2-tensor
in the tangent space of the sphere,
while $s^3$ is a scalar.
To obtain a global description capable of
handling correlations in CMB data across the sky,
it is convenient to work in a spin-weighted basis. 
By definition,
a spin-weighted function $_sf$ of weight $s$
transforms according to
${}_sf'=e^{-s\de} {}_sf$
under a local rotation by $\de$
in the tangent space of the sphere.
We define spin-weighted Stokes parameters
$s_{(\pm 2)}=s^1\mp i s^2$
of spin-weight $\pm2$
and $s_{(0)}=s^3$ of spin-weight 0,
and we adopt the spin-weighted basis 
in which the Stokes vector becomes
$\mbf{s} = (s_{(+2)},s_{(0)},s_{(-2)})^T$.
With these definitions,
a global description can be achieved 
by decomposing various quantities of interest 
in terms of spin-weighted spherical harmonics 
$\sylm{s}{lm}(\hat n)$
\cite{swsh1,swsh2}.
The $\sylm{s}{lm}(\hat n)$
can be viewed as the generalization 
of the usual spherical harmonics to tensors 
in the tangent space of the sphere,
with integer indices restricted 
by $l\geq |s|$ and $m=-l,\ldots,l$.
For a fixed spin weight $s$,
the $\sylm{s}{lm}(\hat n)$
form a complete orthonormal set 
of spin-$s$ functions on the sphere.

The CMB temperature $T$ and the Stokes parameter $s^3=s_{(0)}$ 
are scalars on the sphere 
and can be decomposed into
the usual spherical harmonics $Y_{lm}\equiv \sylm{0}{lm}$,
while the Stokes parameters $s_{(\pm 2)}$ 
are combinations of harmonics with spin-weight $\pm 2$:
\bea
&
T = \sum_{lm} a_{(T)lm}\ \sylm{0}{lm} ,
\quad
s_{(0)} = \sum_{lm} a_{(V)lm}\ \sylm{0}{lm},
\nonumber\\
& s_{(\pm 2)} = \sum_{lm} (a_{(E)lm} \pm i a_{(B)lm})\ \sylm{\pm2}{lm}.
\label{amps}
\eea
Here,
each amplitude obeys 
$a_{(X)lm}^* = (-1)^m a_{(X)l -\! m}$
with $X=T,E,B,V$.
The notation $E$ and $B$ arises from the parity properties
of the amplitudes, 
which mimics those of the electric and magnetic fields,
while the notation $V$ arises from the Stokes usage
for circular polarization.
The above decompositions are convenient 
since general considerations predict 
the CMB has no $V$-type (circular) polarization 
and significant nonzero cross correlations 
only between the $T$ and $E$ amplitudes
when reionization or other foreground effects are neglected
\cite{cmbrev}.
Typically,
CMB observations of temperature and polarization 
are expressed as estimates of power spectra and correlations 
via the coefficients
$C^{X_1X_2}_l \equiv 
\frac{1}{2l+1}\sum_m \vev{a^*_{(X_1)lm}a_{(X_2)lm}}$.
In the absence of Lorentz violation,
a $TT$ component,
a small $EE$ component, 
and a $TE$ correlation are predicted,
consistent with existing data at present sensitivities.
Even smaller $BB$ modes without $TB$ correlations are also expected,
but confirmation of this lies beyond current observational reach.

The Stokes vector $\mbf\vs$ determining the rotation axis
for birefringence
can also be decomposed in the spin-weighted basis,
$\mbf\vs = (\vs_{(+2)},\vs_{(0)},\vs_{(-2)})^T$.
This gives
\beq
\Si=2E\left(
\begin{array}{ccc}
\vs_{(0)}         & -\vs_{(+2)} & 0 \\
-\half\vs_{(-2)}  & 0           & \half\vs_{(+2)} \\
0                 & \vs_{(-2)}  & -\vs_{(0)}
\end{array}\right) .
\label{gen}
\eeq
The components $\vs_{(s)}$ can be written explicitly 
in terms of the coefficients for Lorentz violation
in Eqs.\ \rf{kfs}.
However,
it is convenient here to expand in spin-weighted spherical harmonics.
Some calculation yields 
\bea
&\vs_{(0)} = \sum_{d}\sum_{lm} 
E^{d-4} \klm{d}{V}{lm}\ \sylm{0}{lm} , 
\nonumber\\
&\vs_{(\pm 2)} = \sum_{d}\sum_{lm} 
E^{d-4} ( \klm{d}{E}{lm} \pm i \klm{d}{B}{lm} )\ \sylm{\pm 2}{lm} .
\label{kamps2}
\eea
Here, 
$d$ is odd for $\vs_{(0)}$,
$d$ is even for $\vs_{(\pm 2)}$,
$l \leq d-2$ for both,
and $l\geq 2$ for $\vs_{(\pm 2)}$.
It follows that CMB Lorentz violations 
separate into three categories $E$, $B$, $V$
according to the operator dimension $d$ 
and its P and CPT properties.
The coefficients for Lorentz violation 
$\klm{d}{E}{lm}$,
$\klm{d}{B}{lm}$,
$\klm{d}{V}{lm}$
are constants of dimension $E^{4-d}$.
The $E$, $B$ effects preserve CPT,
while the $V$ effects violate it. 
  
A complete analysis of available CMB polarization data 
\cite{boom_pol,dasi_pol,capmap_pol,cbi_pol,wmap_pol}
is challenging because searching for Lorentz violation requires 
careful treatment of the frequency dependences.
We avoid these complications here by focusing 
on results from the BOOMERANG (B03) experiment
\cite{boom_pol},
which performed polarimetry 
in a single relatively narrow high-frequency band 
at approximately 145 GHz.
The effects grow roughly as $E^{d-3}$,
so inclusion of other lower-frequency results 
may reduce errors but is unlikely
to change sensitivites drastically.
We match to the B03 data
by comparing published values of $C_l$
\cite{boom_pol}
with those expected from nonzero birefringence.
We assume conventional initial $C_l$,
with nonzero $C^{TT}_l$, $C^{TE}_l$, and $C^{EE}_l$ only,
calculated using available software 
\cite{cmbfast}.
The $TT$ data are unaffected by birefringence
and can be disregarded here.
Including them 
and varying the underlying cosmology or the initial $C_l$
to find the joint best-fit cosmological parameters
and Lorentz-violating coefficients is expected 
to yield similar results 
because the $TT$ data dominate the statistics
and our initial $C_l$ are consistent with other larger datasets.
The initial $C_l$ are used to generate polarization maps of the sky.
For chosen values of coefficients for Lorentz violation,
the maps are propagated numerically via Eq.\ \rf{stokesrot}
to the present epoch,
and the $C_l$ predicted today are extracted.
For simplicity,
we consider one nonzero coefficient at a time,
although in principle any combination of coefficients
may exist in nature.
The theoretical $C_l$ are binned to match the reported B03 values for 
$C^{TE}_l$, 
$C^{TB}_l$, 
$C^{EE}_l$, 
$C^{BB}_l$,
and a $\ch^2$ distribution is constructed,
$\ch^2 = \sum_{\mbox{\scriptsize bins}} 
(C_{\rm B03}-C_{\rm theory})^2/(\si_{\rm B03}^2+\si_{\rm theory}^2)$.

\begin{table}
\begin{tabular}{ccc}
\hline
Coefficient & Value & $\chi^2/$d.o.f. \\
\hline
$\klm{3}{V}{00}$& 
$(12\pm7) \times 
10^{-43} \mbox{ GeV}$
& 1.2 
\\
$\klm{3}{V}{10}$& 
$ \pm(3\pm1) \times 
10^{-42} \mbox{ GeV}$
& 1.2 
\\
$\klm{3}{V}{11}$& 
$ \pm(21^{+7}_{-9}) \times 
10^{-43} \mbox{ GeV}$
& 1.2
\\
$\klm{4}{E}{20}$& 
$ \pm(17^{+7}_{-9}) \times 
10^{-31}$
& 1.2 
\\
$\klm{4}{B}{20}$& 
$ \pm(17^{+7}_{-9}) \times 
10^{-31}$
& 1.2 
\\
$\klm{5}{V}{00}$& 
$(3\pm2) \times 
10^{-20} \mbox{ GeV}^{-1}$
& 1.2
\\
$\klm{5}{V}{10}$& 
$ (8^{+2}_{-3}) \times 
10^{-20} \mbox{ GeV}^{-1}$
& 1.2 
\\
&
$ -(8^{+3}_{-4}) \times 
10^{-20} \mbox{ GeV}^{-1}$
& 1.2 
\\
$\klm{5}{V}{20}$& 
$-(10\pm3) \times 
10^{-20} \mbox{ GeV}^{-1}$
& 1.1 
\\
$\klm{5}{V}{30}$& 
$ (8^{+3}_{-4}) \times 
10^{-20} \mbox{ GeV}^{-1}$
& 1.2
\\
&
$ -(8\pm3) \times 
10^{-20} \mbox{ GeV}^{-1}$
& 1.2
\\
$\klm{6}{E}{20}$& 
$ \pm(11^{+4}_{-5}) \times 
10^{-10} \mbox{ GeV}^{-2}$
& 1.2 
\\
$\klm{6}{E}{30}$& 
$ \pm(11^{+5}_{-6}) \times 
10^{-10} \mbox{ GeV}^{-2}$
& 1.2 
\\
$\klm{6}{E}{40}$& 
$ \pm(11^{+5}_{-6}) \times 
10^{-10} \mbox{ GeV}^{-2}$
& 1.2 
\\
\hline
\end{tabular}
\caption{Sample measured 1$\si$ values of coefficients
showing $\chi^2$ per degree of freedom. Each fit is 
performed independently.}  
\end{table}

Figure 1 shows our estimated likelihoods for several types 
of Lorentz violations.
The figure reveals that 
at the $1\si$ level the B03 data prefer nonzero values 
for all coefficients for Lorentz violation
but are consistent with no violations at $2\si$.
Note that for $d>3$ 
the comparatively high B03 frequency leads to somewhat 
tighter constraints than our estimates in Table I,
demonstrating the advantage of higher-energy studies.
For each independent fit,
the preferred values and 1$\si$ ranges of the coefficients
are listed in Table II.
Except for one special case,
all coefficients for Lorentz violation cause
either frequency- or direction-dependent polarization rotations,
resulting in complicated changes in polarization over the sky.
Only coefficients with $d=3$ produce 
frequency-independent effects,
and only the single special coefficient $\klm{3}{V}{00}$
produces polarization rotations 
that are also uniform over the entire sky.
A recent study of this special case \cite{feng}
found that B03 and other CMB data 
favor a small nonzero rotation angle of $6^\circ\pm 4^\circ$,
which in the present context is equivalent to the value
$\klm{3}{V}{00} \simeq (6\pm 4) \times 10^{-43}$ GeV
and is compatible with the result in Table II.
At the 95\% confidence level, 
we obtain an upper limit of 
$\klm{3}{V}{00} \lsim 26 \times 10^{-43}$ GeV.
This is consistent with the constraint
$\klm{3}{V}{00} \lsim 40 \times 10^{-43}$ GeV
obtained from radio-galaxy polarimetry
\cite{cfj}.

Table II also includes various results for 
frequency- and direction-dependent birefringence effects.
We find that 2$\si$ constraints
on the coefficients $\klm{3}{V}{10}$ and $\klm{3}{V}{11}$,
which control anisotropic Lorentz violations for $d=3$,
lie at the level of $10^{-42}$ GeV.
Violations involving operators with $d=4$
are constrained to the 2$\si$ level of $10^{-30}$.
This limit is consistent with the existing partial constraints 
on these coefficients of approximately $10^{-32}$
obtained from spectropolarimetry of galaxies 
and of approximately $10^{-37}$ obtained from gamma-ray bursts 
\cite{km}.
However,
the point-source nature of these previous results means that,
while extremely sensitive,
they only cover a limited portion of the coefficient space.
Among all coefficients with $d=5,6$,
only $\klm{5}{V}{00}$ is direction independent.
Our $2\si$ constraint on this coefficient is consistent 
with studies of its effects in other contexts \cite{k500}.
For the direction-dependent coefficients with $d=5$ and $d=6$
given in Table II,
the measurements listed are the first obtained.

Overall,
our results demonstrate that studies of the CMB polarization 
offer broad sensitivity to possible effects from all 
coefficients for Lorentz violation in electrodynamics.
While incorporation of additional available data 
is unlikely to increase significantly the net sensitivity,
other CMB experiments may provide tests
of the robustness of the $1\si$ birefringent signals 
and determine whether they could be indicative 
of systematic effects or more conventional phenomena
such as foregrounds.
If the signals persist,
existing and future high-resolution polarimetric data 
could determine which types of violations are preferred.
Whatever the outcome,
CMB polarimetry provides highly sensitive tests 
of spacetime symmetries
with the potential to reveal signals of fundamental physics.

This work was supported in part
by DOE grant DE-FG02-91ER40661,
by NASA grant NAG3-2194,
and by the Wisconsin Space Grant Consortium.


\vskip -10pt
\begin{thebibliography}{99}

\bibitem{ks}
V.A.\ Kosteleck\'y and S.\ Samuel,
Phys.\ Rev.\ D {\bf 39}, 683 (1989).

\bibitem{lipa}
J.A.\ Lipa
\etal,
Phys.\ Rev.\ Lett.\ {\bf 90}, 060403 (2003).

\bibitem{muller}
H.\ M\"uller
\etal,
Phys.\ Rev.\ Lett.\ {\bf 91}, 020401 (2003).

\bibitem{stanwix}
P.L.\ Stanwix 
\etal,
Phys.\ Rev.\ Lett.\ {\bf 95}, 040404 (2005).  

\bibitem{herrmann}
S.\ Herrmann
\etal,
Phys.\ Rev.\ Lett.\ {\bf 95}, 150401 (2005).  

\bibitem{antonini}
P.\ Antonini 
\etal,
Phys.\ Rev.\ A {\bf 71}, 050101 (2005).  

\bibitem{stanwix2}
P.L.\ Stanwix
\etal,
Phys.\ Rev.\ D {\bf 74}, 081101 (2006).  

\bibitem{cfj}
S.M.\ Carroll
\etal,
Phys.\ Rev.\ D {\bf 41}, 1231 (1990).

\bibitem{km}
V.A.\ Kosteleck\'y and M.\ Mewes,
Phys.\ Rev.\ Lett.\ {\bf 87}, 251304 (2001);
Phys.\ Rev.\ D {\bf 66}, 056005 (2002);
Phys.\ Rev.\ Lett.\ {\bf 97}, 140401 (2006).  

\bibitem{bear}
D.\ Bear
\etal, 
Phys.\ Rev.\ Lett.\ {\bf 85}, 5038 (2000);
M.A.\ Humphrey 
\etal, 
Phys.\ Rev.\ A {\bf 68}, 063807 (2003);
F.\ Can\`e 
\etal, 
Phys.\ Rev.\ Lett.\ {\bf 93}, 230801 (2004).

\bibitem{wolf}
P.\ Wolf
\etal,
Phys.\ Rev.\ Lett.\ {\bf 96}, 060801 (2006).  

\bibitem{heckel}
B.R.\ Heckel
\etal, 
Phys.\ Rev.\ Lett.\ {\bf 97}, 021603 (2006).  

\bibitem{sme}
D.\ Colladay and V.A.\ Kosteleck\'y, 
Phys.\ Rev.\ D {\bf 55}, 6760 (1997);
Phys.\ Rev.\ D {\bf 58}, 116002 (1998);
V.A.\ Kosteleck\'y,
Phys.\ Rev.\ D {\bf 69}, 105009 (2004).

\bibitem{kp}
V.A.\ Kosteleck\'y and R.\ Potting,
Nucl.\ Phys.\ B {\bf 359}, 545 (1991);
Phys.\ Rev.\ D {\bf 51}, 3923 (1995).

\bibitem{ng}
R.\ Bluhm and V.A.\ Kosteleck\'y,
Phys.\ Rev.\ D {\bf 71}, 065008 (2005);
B.\ Altschul and V.A.\ Kosteleck\'y,
Phys.\ Lett.\ B {\bf 628}, 106 (2005).

\bibitem{boom_pol}
T.E.\ Montroy 
\etal,
Ap.\ J.\ {\bf 647}, 813 (2006);
F.\ Piacentini 
\etal,
Ap.\ J.\ {\bf 647}, 833 (2006).

\bibitem{dasi_pol}
E.M.\ Leitch
\etal,
Ap.\ J.\ {\bf 624}, 10 (2005).

\bibitem{capmap_pol}
D.\ Barkats
\etal,
Ap.\ J.\ {\bf 619}, L127 (2005).

\bibitem{cbi_pol}
A.C.S.\ Readnead 
\etal,
Science {\bf 306}, 836 (2004).

\bibitem{wmap_pol}
L.\ Page 
\etal,
astro-ph/0603450.

\bibitem{feng}
B.\ Feng
\etal,
Phys.\ Rev.\ Lett.\ {\bf 96}, 221302 (2006).

\bibitem{cmb}
A.\ Lue
\etal,
Phys.\ Rev.\ Lett.\ {\bf 83}, 1506 (1999);
K.R.S.\ Balaji 
\etal,
JCAP {\bf 0312}, 008 (2003);
G.-C. Liu
\etal,
Phys.\ Rev.\ Lett.\ {\bf 97}, 161303 (2006).

\bibitem{swsh1}
E.T.\ Newman and R.\ Penrose,
J.\ Math.\ Phys.\ {\bf 7}, 863 (1966).

\bibitem{swsh2}
J.N.\ Goldberg 
\etal,
J.\ Math.\ Phys.\ {\bf 8}, 2155 (1967).

\bibitem{cmbrev}
W.\ Hu and M.\ White,
New Astron.\ {\bf 2}, 323 (1997).

\bibitem{cmbfast}
U.\ Seljak and M.\ Zaldarriaga,
Ap.\ J.\ {\bf 469}, 437 (1996).

\bibitem{k500}
T.A.\ Jacobson
\etal,
Phys.\ Rev.\ Lett.\ {\bf 93}, 021101 (2004);
R.C.\ Myers and M.\ Pospelov,
{\it ibid.}
{\bf 90}, 211601 (2003).

\end{thebibliography}
\end{document}